\documentclass[twocolumn,showpacs,preprintnumbers,amsmath,amssymb,pra]{revtex4}

\usepackage{graphicx}
\usepackage{dcolumn}
\usepackage{bm}

\begin{document}

\preprint{}

\title{Effects of a localized beam on the dynamics of excitable cavity solitons}

\author{Adrian Jacobo}
\email{jacobo@ifisc.uib-csic.es}
\author{Dami\`a Gomila}
\email{damia@ifisc.uib-csic.es}
\author{Manuel A. Mat\'ias}
\email{Manuel.Matias@ifisc.uib-csic.es}
\author{Pere Colet}
\email{pere@ifisc.uib-csic.es}
 \affiliation{IFISC, Instituto de F\'{\i}sica Interdisciplinar y Sistemas Complejos (CSIC-UIB), E-07122 Palma de Mallorca, Spain}
\homepage{http://ifisc.uib-csic.es}

\date{\today}

\begin{abstract}

We study the dynamical behavior of dissipative solitons in an optical cavity filled
with a Kerr medium when a localized beam is applied on top of the homogeneous pumping.
In particular, we report on the excitability regime that cavity solitons exhibits
which is emergent property since the system is not locally excitable. The resulting scenario
differs in an important way from the case of a purely homogeneous pump and now two
different excitable regimes, both Class I, are shown.
The whole scenario is presented and discussed, showing that it is organized by three
codimension-$2$ points. Moreover, the localized beam can be used to control important
features, such as the excitable threshold, improving the possibilities for the experimental
observation of this phenomenon.
\end{abstract}


\maketitle

\section{Introduction} \label{sec:intro}

Dissipative solitons (also known as localized structures)
are states in extended media that consist of one (or more)
regions in one state surrounded by a region in a qualitatively different
state (in the following this {\it surrounding\/} state is an area
in a stable stationary state). These structures were first suggested in
Refs.~\cite{KogaKura,KernerOsipov} and then described in a variety
of systems, such as chemical reactions \cite{VanagEpstein07},
semiconductors \cite{Niedernostheide}, granular media \cite{Umbanhowar},
binary-fluid convection \cite{Niemela,Batiste}, vegetation patterns \cite{Lejeuneveg},
and also in nonlinear optical cavities where they are usually referred as Cavity Solitons (CS)
\cite{FirthOPN,Lugiato03,Rosanov,barland,Ackemann08} (see
Ref.~\cite{Akhmediev,TTKChaos07,Akh2} for recent surveys). Their potential in optical
storage and processing of information has been stressed \cite{coullet04}. In this work we shall consider
solitons that appear in a subcritical bifurcation \cite{Fauve90,Tlidi94,Rosanov}.

In general, dissipative solitons may develop a number of instabilities like start moving,
breathing, or oscillating. In the latter case, they would oscillate in time
while remaining stationary in space, like the oscillons (oscillating localized
structures) found in a vibrated layer of sand \cite{Umbanhowar}. The
occurrence of these oscillons in autonomous systems has
been reported both in optical \cite{firth96,FirthJOSAB} and chemical systems
\cite{oscillonchem}. It has been shown that they can become
unstable leading to excitable solitons in systems for which the local
dynamics is not excitable \cite{damia05,hompump}. In this case excitability appears as
an emergent property arising from the spatial dependence, which allows for the
formation of these structures. In particular, for solitons arising in uniformly pumped Kerr
cavities, excitability is mediated by a saddle-loop (homoclinic) (SL) bifurcation, and
it is characterized by a large excitability threshold and
by occurring at any point of space that is properly excited \cite{damia05,hompump}.

Since CS excitability emerges from the spatial dependence it is interesting to study
the effect of breaking the translational symmetry on the excitable dynamics.
In optical systems this can be easily done by applying a (small amplitude)
localized beam on top of the homogeneous pump. Addressing beams are typically used already
to create CS by applying a transient perturbation. Here we analyze the dynamics of CS in a Kerr cavity where we apply a permanent addressing beam. On one
hand, this pump allows to control the place where a CS appears. On the other hand, the
system remains excitable and a new route, mediated by a Saddle-Node on an Invariant Circle (SNIC)
bifurcation, appears. It is characterized by the fact that the excitable threshold
is fully tunable, as it scales with the proximity to the bifurcation.

This paper is organized as follows. The model and overall
dynamical behavior exhibited by the system in parameter space are described in
Sect. \ref{sec:model} and \ref{sec:overview}. Sect.~\ref{sec:snic}
addresses the instability exhibited by CS through a SNIC bifurcation.
Sect.~\ref{sec:excitab} discusses the excitable routes found in this system.
Sect.~\ref{sec:cusp} and \ref{sec:snsl} discuss the codimension-$2$ points that
organize the overall scenario. Finally, concluding remarks are given in Sect.~\ref{sec:conclu}.

\section{Model} \label{sec:model}

A prototype model describing an optical cavity filled with a nonlinear
Kerr medium is the one introduced by Lugiato and Lefever \cite{lugiato87}
with the goal of studying pattern formation in this optical system. Later
studies showed that this equation also exhibits CS in
some parameter regimes \cite{firth96,firthlord96}.
The model describes the dynamics of the slowly varying
amplitude of the electromagnetic field $E(\vec{x},t)$ in the paraxial and mean
field approximations ($\vec{x}=(x,y)$ is the plane transverse to the
propagation direction $z$ on which the slow dynamics takes place). The dynamics
of the field is given by:
\begin{eqnarray}
\frac{\partial E}{\partial t}=-(1+i \theta)E+i
\nabla^2E+E_I(\vec{x})+i |E^2|E \label{kerreq}
\end{eqnarray}
where $\nabla^2=\partial^2/\partial x^2+\partial^2/\partial y^2$.

The first term in the right hand side describes cavity losses, $E_I(\vec{x})$ is the input field (pump), $\theta$ is the cavity detuning with respect to $E_I$, and the sign of the cubic term represents
the self-focusing case. Notice that in the absence of losses and of an
input field, the field can be rescaled to $E \rightarrow  E e^{i\theta t}$ to
remove the detuning term and  Eq. (\ref{kerreq}) becomes the Nonlinear Schr\"odinger
Equation (NLSE). It is well documented that in this case, and in two spatial
dimensions, the NLSE exhibits the so called collapse regime \cite{collapse}, in
which energy accumulates at a point of space. Collapse is prevented in Eq.
(\ref{kerreq}) by the cavity losses leading to stable CS.

For spatially homogeneous pump $E_I(\vec{x})=E_0$, Eq. (\ref{kerreq}) has a
homogeneous steady state solution given implicitly by $E_0=E_s[1+i(\theta-I_s)]$,
where $I_s=|E_s|^2$ \cite{lugiato87}. This solution is stable for low pump strength,
that is for $I_s<1$. At $I_s=1$, the so-called modulation instability (MI) point,
the homogeneous solution becomes unstable and extended patterns appear subcritically.
The patterns arising at MI are typically oscillatory and increasing
the pump they undergo further instabilities which eventually lead to optical turbulence
\cite{pra68, pre76}. Static hexagonal patterns can be found subcritically, that is,
decreasing the pump value below the MI point. CS appear in the region of bistability
between the homogeneous solution and the pattern. In fact there are two CS that appear
through a saddle-node (fold) bifurcation, the one with larger amplitude (upper-branch CS) is
stable at least for some parameter range, while the one with smaller amplitude
(middle-branch CS) is always unstable. Early studies already identified that
the upper branch CS may undergo a Hopf bifurcation leading to a oscillatory behavior
\cite{firth96}. The oscillatory instabilities, as well as azimuthal instabilities, were fully
characterized later \cite{FirthJOSAB}. As one moves in parameter space away from the
Hopf bifurcation, the CS oscillation amplitude grows, and finally the
limit cycle touches the middle-branch CS in a saddle-loop bifurcation which leads to a
regime of excitable dissipative structures \cite{damia05,hompump}.

Here, we consider a pump beam of the form
\begin{equation}
E_I(r)=E_0+H\,\exp(-r^2/r_0^2)
\label{addressing}
\end{equation}
where $E_0$ is a homogeneous field, assumed real,
$H$ the height of the localized Gaussian perturbation, $r^2=x^2+y^2$,
while $r_0=1$.
For convenience,  we write the height of the Gaussian beam as,
\begin{equation}
H=\sqrt{(I_s+I_{sh})\,[1+(\theta-I_s-I_{sh})^2]}-E_0,
\label{height}
\end{equation}
where  $I_s$ is the background intracavity intensity (due to $E_0$) and
$I_s+I_{sh}$ corresponds to the intracavity field intensity of a cavity
driven by an homogeneous field with a amplitude equal to one at the top
of the Gaussian beam, $E_I=E_0+H$. This directly relates the height of the
Gaussian beam $H$ with the equivalent intracavity intensity for a homogeneous pump.
Notice that for $I_{sh}=0$ the pump beam becomes homogeneous, $H(I_{sh}=0)=0$,
With the inclusion of the localized pump beam the system has now three independent
control parameters which for convenience take as the background intensity, $I_s$,
the detuning $\theta$ and $I_{sh}$.

The numerical methods used to study this system are detailed in the Appendix of
Ref.~\cite{hompump}. Eq.~(\ref{kerreq}), with the applied pump
(\ref{addressing}), has been solved numerically using a pseudospectral method,
where the linear terms are integrated exactly in Fourier space, while the
nonlinear ones are integrated using a second order in time approximation.
Periodic boundary conditions in a square lattice of size $512\times 512$ points
were used. The stability of steady CS has been studied using the
semi-analytical method discussed in the above mentioned Appendix, using the
radial version of Eq.~(\ref{kerreq}), that simplifies the study taking into
account the axisymmetric nature of the solutions.

\section{Overview of the behavior of the system}
\label{sec:overview}

\begin{figure}
\includegraphics[width=0.4\textwidth]{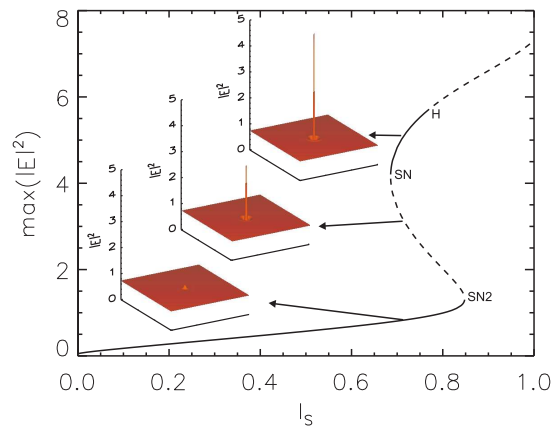}
\caption{(Color online) Bifurcation diagram, $max(I)$ vs $I_s$,
for a CS for $I_{sh}=0.3$, $\theta=1.25$.
Solid lines represent stable solutions and dashed lines unstable
ones. The insets show the transverse profile of the solutions.}
\label{bifis}
\end{figure}

One of the main consequences of the application of a localized pump is
the breaking of the translational symmetry of Eq.~(\ref{kerreq}).
Solutions are now pinned in the region in which the Gaussian pump is
applied. This also affects the transverse profile of the solutions, in particular
the fundamental solution is no longer spatially homogeneous but it exhibits a bump
as illustrated by the lower inset in Fig. 1.

To understand better the effects of the application of a localized pump, a diagram like the one shown in Fig.~2 of Ref.~\cite{hompump}, that represents the maximum intensity of the
transverse field as a function of $I_s$ is shown in Fig.~\ref{bifis},
namely for $I_{sh}=0.3$ and $\theta=1.25$. The diagram, with three branches, looks qualitatively
equivalent to the case of a homogeneous pump and operations such as switching on and off the CS can be performed in a similar way.
For example, with the system at the fundamental solution, the upper branch CS can be switched on by applying an additional transient localized beam or
equivalently by temporarily increasing $I_{sh}$

\begin{figure}
\includegraphics[width=0.49\textwidth]{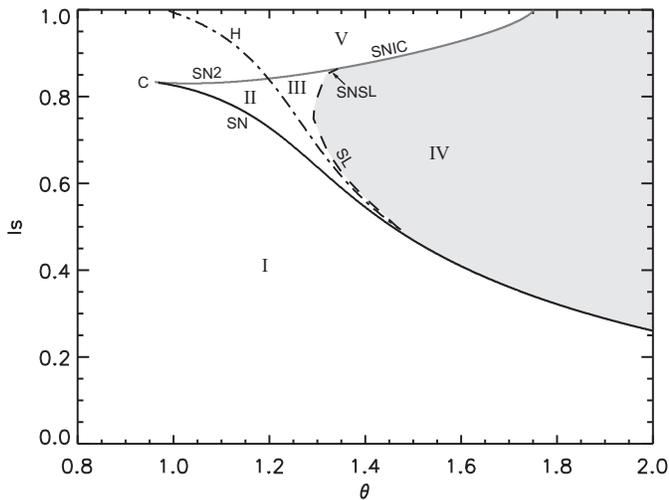}
\caption{Two-parameter $I_{s}$ vs. $\theta$ phase diagram for $I_{sh}=0.3$.
Bifurcation lines are: SN (Saddle-Node); H (Hopf); SL (Saddle-Loop); SNIC
(Saddle-Node on the Invariant Circle); SN2 (Saddle-Node off invariant cycle).
Regions delimited by bifurcation lines are as follows.
I: only the fundamental solution is stable; II: stationary stable CS
coexisting with the fundamental one; III: oscillating CS (or oscillons), coexisting
with the fundamental solution; IV: excitable region; V: oscillating CS (with no
other coexisting solution).}
\label{phasediag_0.3}
\end{figure}

A relevant difference is that for an homogeneous pump the lowest branch
(homogeneous solution) extends until the MI at $I_s=1$, while in the case considered
here the fundamental solution merges with the middle branch CS before the MI, in a
saddle-node bifurcation (that happens at $I_s=0.8479$, SN point in Fig.~\ref{bifis}).
To understand qualitatively this phenomenon one has to take into account that
the homogeneous pump case has many symmetries, some of which are broken when a
localized pump is applied. In technical parlance one says that the bifurcation
has become imperfect (see, e.g., \cite{strogatz}, with the consequence that a
gap in $I_s$ appears making the lower branch disconnected in two branches (the
right part of the branch is not plotted in Fig.~\ref{bifis} and correspond to
solutions unstable to extended patterns).

Exploring now the upper branches, in Fig.~\ref{bifis} past the saddle-node
bifurcation at $I_s=0.6857$ (SN point), a pair of stationary (stable, upper branch, and unstable, middle branch) localized solutions in the form of CS are found.
In this parameter region, these structures are not essentially different to
the solutions found in the homogeneous case \cite{hompump}. Increasing
$I_{s}$ the stable high-amplitude CS undergoes a Hopf bifurcation.

\begin{figure}
\includegraphics[width=0.49\textwidth]{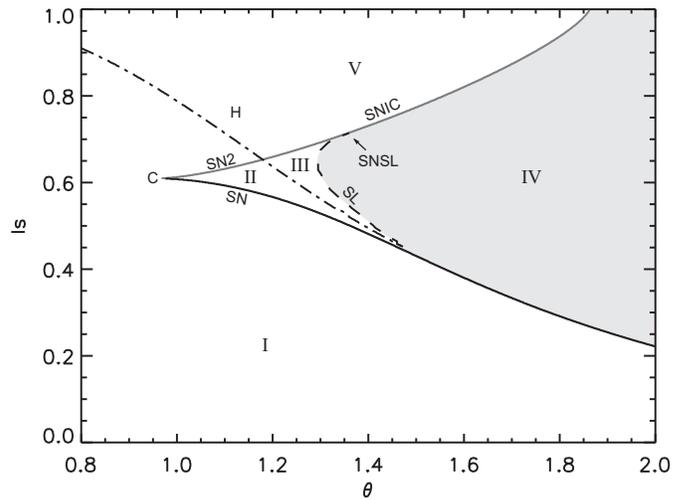}
\caption{Two-parameter $I_{s}$ vs. $\theta$ phase diagram for $I_{sh}=0.7$.
Line and region labeling as in Fig.~\ref{phasediag_0.3}.
}
\label{phasediag_0.7}
\end{figure}

Overall, the scenario found for a localized pump is much more involved than in the
case of the homogeneous case, and more types of behavior are found in this case.
Fig.~\ref{phasediag_0.3} shows a phase diagram for a fixed value of the localized pump $I_{sh}=0.3$.
One can compare this figure with Fig.~1 in Ref.~\cite{hompump}, corresponding to $I_{sh}=0$.
The effect of breaking the translational symmetry would
be to unfold some of the lines at $I_s=1$ (not visible in Fig.~1 of Ref.~\cite{hompump}),
that are degenerate with the MI line, and also
make the SN line end at $I_s<1$ (point C in Fig.~\ref{phasediag_0.3}). Thus, the effect of
a localized pump is to {\it push\/} down the SN2-SNIC line (to be explained later), as is clear
from  Fig.~\ref{phasediag_0.7}, that provides a similar plot for $I_{sh}=0.7$.

Some of the most prominent features of these figures, comparing with the homogeneous pump
case, associated to the appearance of the SNIC line (to be discussed in more detail in
Sec.~\ref{sec:snic}) are that the excitable region, IV in
Figs.~\ref{phasediag_0.3}-\ref{phasediag_0.7}, can have two types of excitable behavior
(see Sec.~\ref{sec:excitab}), both of Class I, as two different transitions to oscillatory
behavior are possible, saddle-loop (SL) and SNIC. In addition, one has a new region, V, in which
one has a single attractor in the system, that is oscillatory, to be distinguished from
region III, in which the system exhibits bistability \footnote{When speaking about
mono and bistability we refer to localized solutions. In general, several extended solutions
(patterns) may also be coexisting}, between the (stationary) fundamental
solution and an oscillatory upper branch CS. These behaviors, and how they are organized by
three codimension-$2$ points will the subject of Sections~\ref{sec:cusp}-\ref{sec:snsl}.

\section{Saddle-node on the invariant circle bifurcation}
\label{sec:snic}

A saddle-node on the invariant circle bifurcation (SNIC), also known as
saddle-node infinite-period (SNIPER) and as saddle-node central homoclinic
bifurcation, is a special case of a saddle-node
bifurcation that occurs inside a limit cycle. Although this
bifurcation is local in (one-dimensional) flows on the circle, it has global
features in higher-dimensional dynamical systems \cite{strogatz}, so it is also
termed local-global or semilocal. In particular, the (stable) manifolds of the
saddle and node fixed points transverse to the center manifold are organized
by an unstable focus inside the limit cycle. At one side of the bifurcation the
system exhibits oscillatory behavior, while at the other side the dynamics of the
system is excitable. This mechanism leading to excitability has been found in
several (zero dimensional) optical systems \cite{CoulDabTred,Giudici97,HuyetSNIC}.

\begin{figure}
\includegraphics[width=0.4\textwidth]{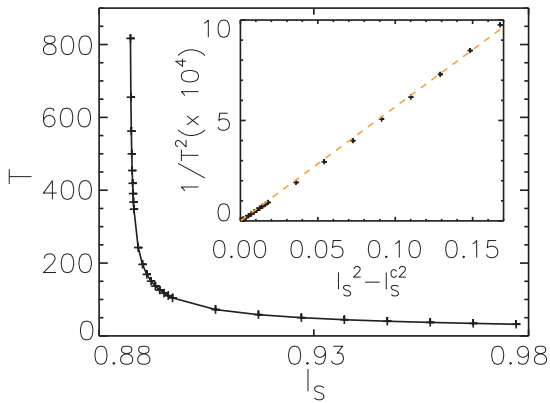}
\caption{Period of the limit cycle $T$ as a function of $I_s$
for $I_{sh}=0.3$ and $\theta=1.45$. Inset: $1/T^2$ vs. $I_s$ close to the
bifurcation point.}
\label{sclaw}
\end{figure}

When approaching the bifurcation from the oscillatory side the period lengthens and
becomes infinite. Quantitatively the period as a function of
a parameter exhibits a inverse square root singular law \cite{strogatz},
\begin{eqnarray}
T&\propto&[I_s^2-(I^c_s)^2]^{-1/2}\ . \label{scalaw}
\end{eqnarray}
This can be used to distinguish the SNIC from other bifurcations leading to
oscillatory behavior (e.g. from the saddle-loop bifurcation with a logarithmic
singular law \cite{damia05,hompump,strogatz}). Fig.~\ref{sclaw} shows
the period of the oscillations as a function of $I_{s}$ obtained by numerical simulations of Eq.~(\ref{kerreq}).
In the inset of Fig.~\ref{sclaw} we plot $1/T^2$ vs. $I_s^2$ close to the
bifurcation point. The linear dependence obtained corroborates the scaling law
and that the transition takes place through a SNIC bifurcation.

\begin{figure}
\includegraphics[width=0.49\textwidth]{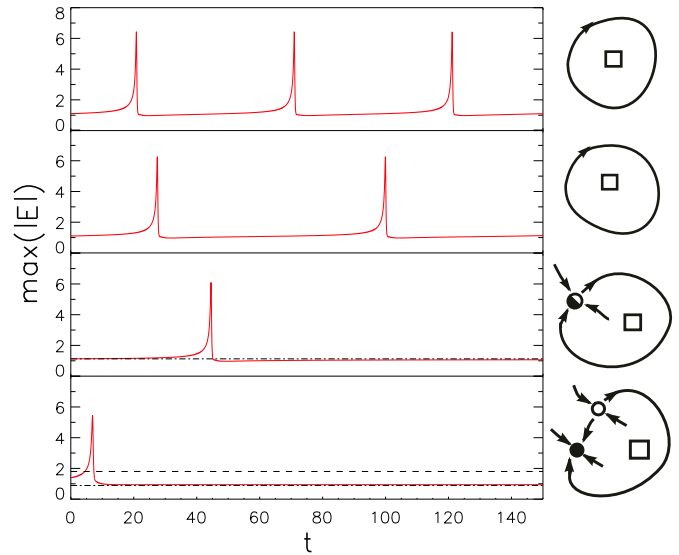}
\caption{(Color Online) Time evolution of the CS amplitude for $I_{sh}=0.3$, $\theta=1.45$ and
decreasing values of $I_s$ . From top to bottom, $I_s= 0.927,
0.907, 0.8871, 0.8$. In the bottom panel the dashed line corresponds to the amplitude of the unstable CS (saddle) while dot-dashed line corresponds to the stable fundamental solution. These two solutions coincide when the SNIC bifurcation takes place (third panel). The sketches on the right illustrate the phase space dynamics.}
\label{sketch_snic}
\end{figure}

The overall route exhibited by the system along a vertical cut in Fig.~\ref{phasediag_0.3} at $\theta=1.45$ is illustrated in Fig.~\ref{sketch_snic},
as the parameter $I_s$ is decreased from the top to the bottom of the Figure.
Thus, the figure shows the transition from oscillatory behavior (top
panel) to stationary (fourth panel) through the occurrence of a SNIC bifurcation
(third panel). Lengthening of the period can be seen in the second panel.
The sketches shown in the right column of Fig.~\ref{sketch_snic} illustrate
the structure of the phase space. The validity of this scenario is further reinforced
with the quantitative analysis presented in Section~\ref{sec:snsl}.

\section{Excitability} \label{sec:excitab}

An interesting aspect of this scenario is that in region IV one can have
excitable behavior through two different mechanisms. On the one hand, and similarly
to the behavior analyzed in Refs.~\cite{damia05,hompump}, close enough to the SL
line one has excitability if the fundamental solution is appropriately excited
such that the oscillatory behavior existing beyond the SL is transiently
recreated. The second mechanism takes place close to the SNIC line, where the oscillatory behavior that
is transiently recreated is that of the oscillations in region V. Both excitable
behaviors exhibit a response starting at zero frequency (or infinite period), as both
bifurcations are mediated by a saddle, whose stable manifold is the threshold
beyond which perturbations must be applied to excite the system. In neuroscience
terminology, both excitable behaviors are class (or type) I
\cite{IzhikevichIJBC,IzhikevichDSN}, although there are important differences
between them. The SNIC mediated excitability is easier to observe than the one
associated to a saddle-loop bifurcation for two reasons. First, it occurs in a
broader parameter range due to its square-root scaling law (\ref{scalaw}), with
respect to the SL excitability \footnote{For saddle-loop mediated
excitability the scaling law is logarithmic, implying that the frequency
increases very fast from zero in a very narrow range, and, thus, its class I
features can be easily missed experimentally \cite{IzhikevichDSN}.}.
Second the excitable threshold can be controlled by the intensity of the localized Gaussian beam, that effectively
approaches the fixed point and the saddle in phase space, allowing to reduce the
threshold as much as desired (by approaching the SNIC line). Within region IV one can find
a typical crossover behavior for the threshold, as it increases from zero (at the SNIC line)
to the (finite) value characteristic of the SL bifurcation as one approaches this line.

\begin{figure}
\includegraphics[width=0.49\textwidth]{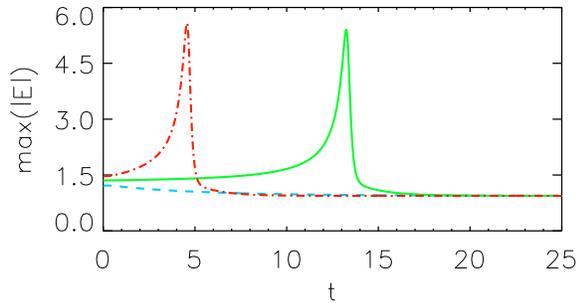}
\caption{(Color online) Evolution of the maximum of the field amplitude after applying a
localized perturbation to the fundamental solution. The perturbation has the shape of the unstable stationary CS (saddle), that is scaled by $0.95$ (blue dashed line); $1.01$ green solid line) and $1.1$ (red dotted line).
Here $I_{sh}=0.3$, $I_s=0.8$, and $\theta=1.45$.} \label{excit3t}
\end{figure}

\begin{figure}
\includegraphics[width=0.49\textwidth]{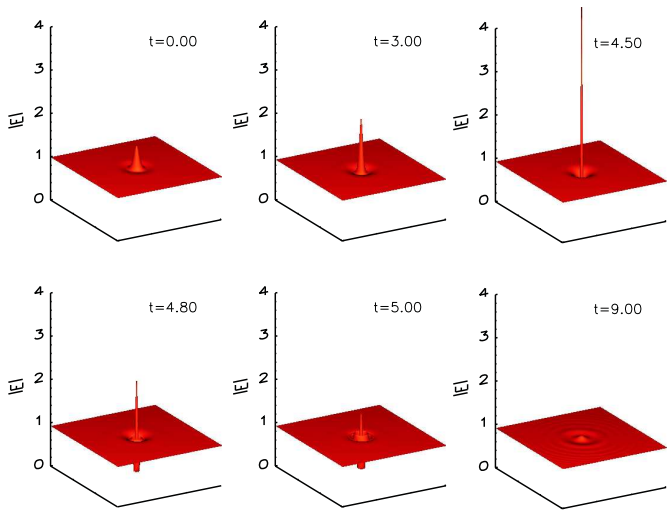}
\caption{(Color online) Transverse profile of $|E|$ at different
times of the dotted line in Fig.~\ref{excit3t}.} \label{excitsnap}
\end{figure}

Fig.~\ref{excit3t} shows the dynamics of the excitable fundamental solution in region IV, namely for the parameters corresponding to the fourth panel in Fig.~\ref{sketch_snic}, upon the application of
different localized perturbations, one below the excitable threshold and two
above. As expected, the perturbation below threshold relaxes directly to the
fundamental solution while the above threshold perturbations elicit first a
large response of the system in the form of an excitable soliton which finally
relaxes to the fundamental solution. The excitable excursion takes place at a later time as the smaller is
the distance to the threshold (a signature of Class I excitability). Finally, the shape of an excitable excursion in
two-dimensional space is shown in Fig.~\ref{excitsnap}.

\section{Cusp codimension-$2$ point} \label{sec:cusp}

\begin{figure}
\includegraphics[width=0.4\textwidth]{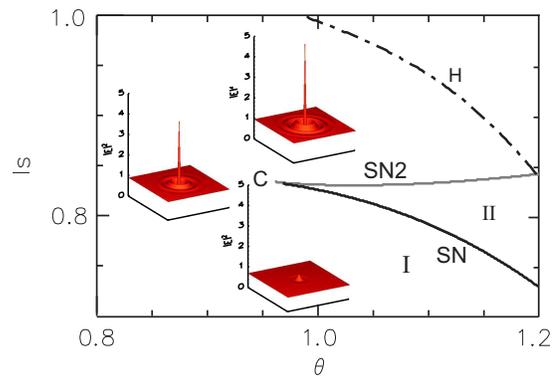}
\caption{(Color online) The fundamental solution for three points in the $(I_s,\theta)$ parameter
space, for $I_{sh}=0.3$, around the Cusp codimension-$2$ point. The coordinates are (bottom
to top): $I_{sh}=0.7, \theta=0.9$; $I_{sh}=0.85, \theta=0.85$, $I_{sh}=0.95, \theta=0.95$.
} \label{aroundcusp}
\end{figure}

In the two-parameter phase diagrams in Figs.~\ref{phasediag_0.3}-\ref{phasediag_0.7}
one can see a point marked with a 'C' that was not found in the homogeneous case
\footnote{The Cusp coordinates are $I_s=0.834, \theta=0.961$ for $I_{sh}=0.3$
(Fig.~\ref{phasediag_0.3}), and $I_s=0.61, \theta=0.968$ for $I_{sh}=0.7$
(Fig.~\ref{phasediag_0.7})}.
This point represents a Cusp codimension-$2$ bifurcation point \cite{Kuznetsov},
namely a point in which two saddle-node curves merge. This cusp point, that involves
only stationary (saddle-node) bifurcations, is also known as the Cusp Catastrophe
\cite{PostonStewartCT}. For parameter values just at the left of the Cusp the bump of
the fundamental solution exhibits a rapid increase. Fig.~\ref{aroundcusp} shows the sharp,
but smooth, change in the shape of the fundamental solution for three parameter values
around the cusp point within region I. Instead, if one is to the right of the 'C' point this
increase cannot be accommodated smoothly and a double fold occurs, such that three branches
appear: two stable, upper and lower branches, and one unstable, middle branch, and thus
bistability makes its appearance. The two folds, codim-$1$, merge critically at the cusp
point, codim-$2$. Decreasing $I_{sh}$ the Cusp moves up towards $I_s=1$, so in the limit of
homogeneous pump it can not be seen due to the presence of the MI instability.
The smooth connection between the fundamental branch and the upper branch exhibiting CS is
an outcome of the symmetry breaking induced by the localized pump which has made the MI bifurcation to
become imperfect.

\section{Saddle-node separatrix-loop codimension-$2$ point} \label{sec:snsl}

\begin{figure}
\includegraphics[width=0.49\textwidth]{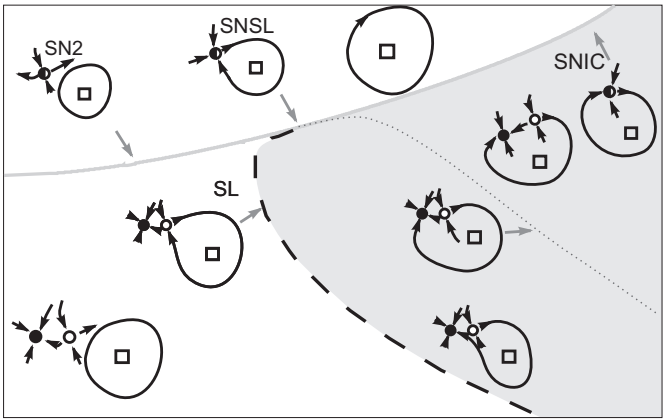}
\caption{Sketch of the parameter space near the SNSL point, showing the unfolding
of bifurcation lines, and adapted to the geometry in
Figs.~\ref{phasediag_0.3}-\ref{phasediag_0.7}.
}
\label{phasesketch}
\end{figure}

The subject of the present section is to discuss the point designated with 'SNSL' (that
stands for Saddle-Node Separatrix Loop \cite{Schecter,IzhikevichIJBC} (also called
Saddle-Node noncentral Homoclinic bifurcation and saddle-node homoclinic orbit bifurcation
\cite{IzhikevichDSN}) in
Figs.~\ref{phasediag_0.3}-\ref{phasediag_0.7}. A SNSL is a local-global codimension-$2$ point
in which a saddle-node bifurcation takes place simultaneously to a saddle-loop,
such that the orbit enters through the noncentral (stable) manifold. The
unfolding of a SNSL point leads to the scenario depicted in Fig.~\ref{phasesketch}.
There is a line of saddle-node bifurcations (in which a pair of stable/unstable fixed points are created)
that at one side of the SNSL is a saddle-node bifurcation off limit cycle (SN2) while
at the other side is a SNIC bifurcation (the saddle-node occurs inside the limit cycle).
A saddle-loop (SL) bifurcation also unfolds from the SNSL point, tangent to the SN2 line. From the other side,
a special curve depicted with a dotted line in Fig.~\ref{phasesketch} should appear (cf.
Fig.~22 in Ref.~\cite{IzhikevichIJBC}). It is not a bifurcation line,
but instead it is a special line in which the approach to the node point is through the most stable
direction, namely the direction transverse to the central manifold at the SNIC/SN2 (and SNSL)
bifurcations (for these reasons we also call this dotted line the 'pseudo-bifurcation' line).
This, in principle nongeneric, curve, that emerges in the unfolding of the SNSL point, is necessary for consistency on the excursions around the SNSL codimension-$2$ point.
At the pseudo-bifurcation the topological cycle is reconstructed, namely the stable fundamental
solution and the middle branch CS (saddle), become points belonging to the circle that latter
will become the limit cycle. Before the pseudo-bifurcation the unstable manifold of the saddle
enters towards the stable solution from the same side than the saddle, while after the
pseudo-bifurcation it enters from the opposite side.


The SNSL point separates two possible ways in which the system can go from
oscillatory region III (where the limit cycle coexists with the stable fundamental solution)
to oscillatory region V (where the fundamental solution does not exist).

\begin{figure}
\includegraphics[width=0.4\textwidth]{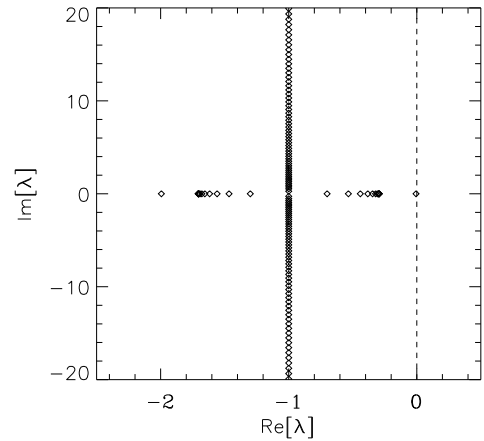}
\caption{Spectrum of the fundamental solution at the SNIC bifurcation for $I_{sh}=0.7$, $I_s=0.707$,
and $\theta=1.34$ (close to the SNSL).}
\label{spectrum}
\end{figure}

\begin{figure}
\includegraphics[width=0.23\textwidth]{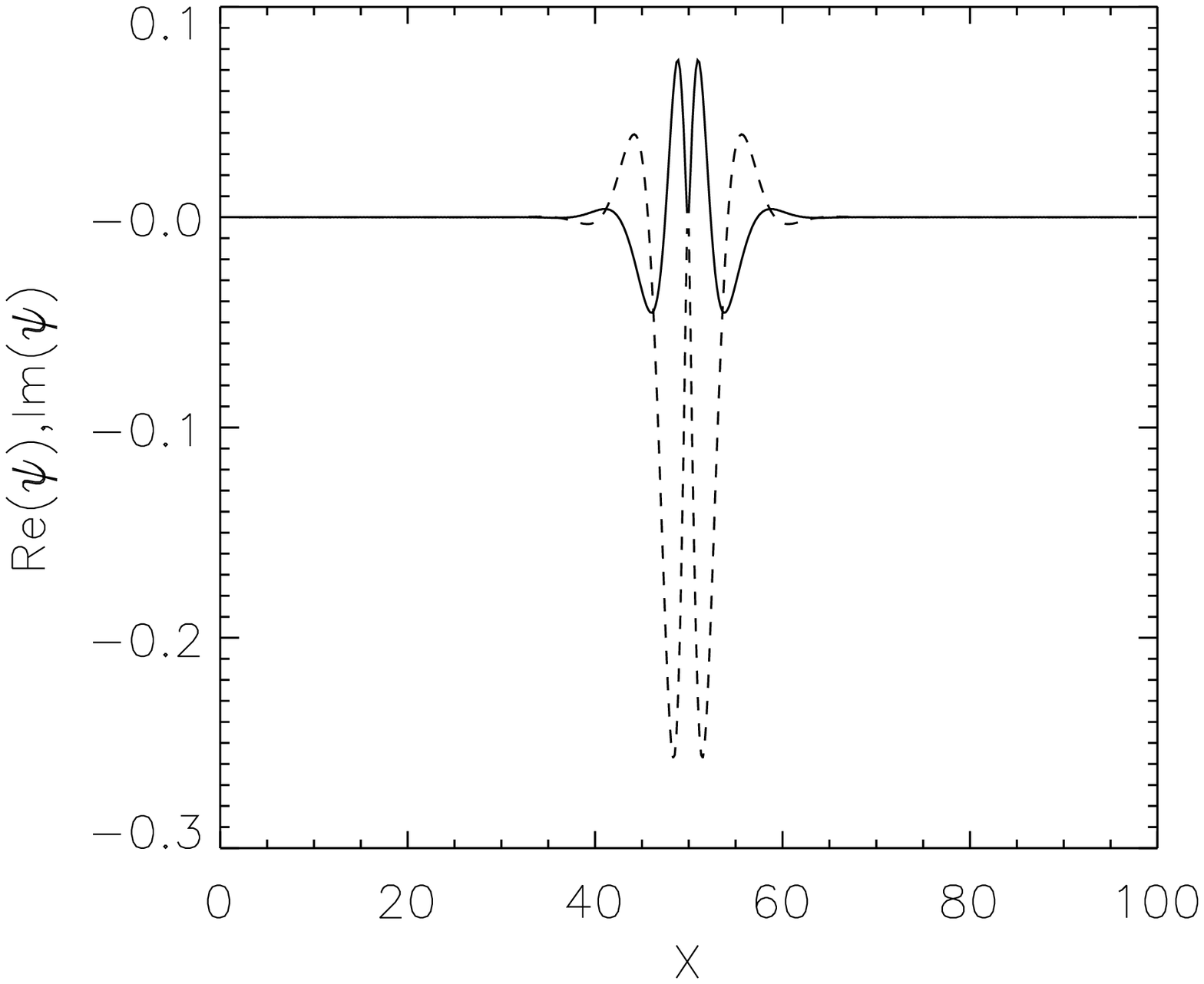}
\includegraphics[width=0.23\textwidth]{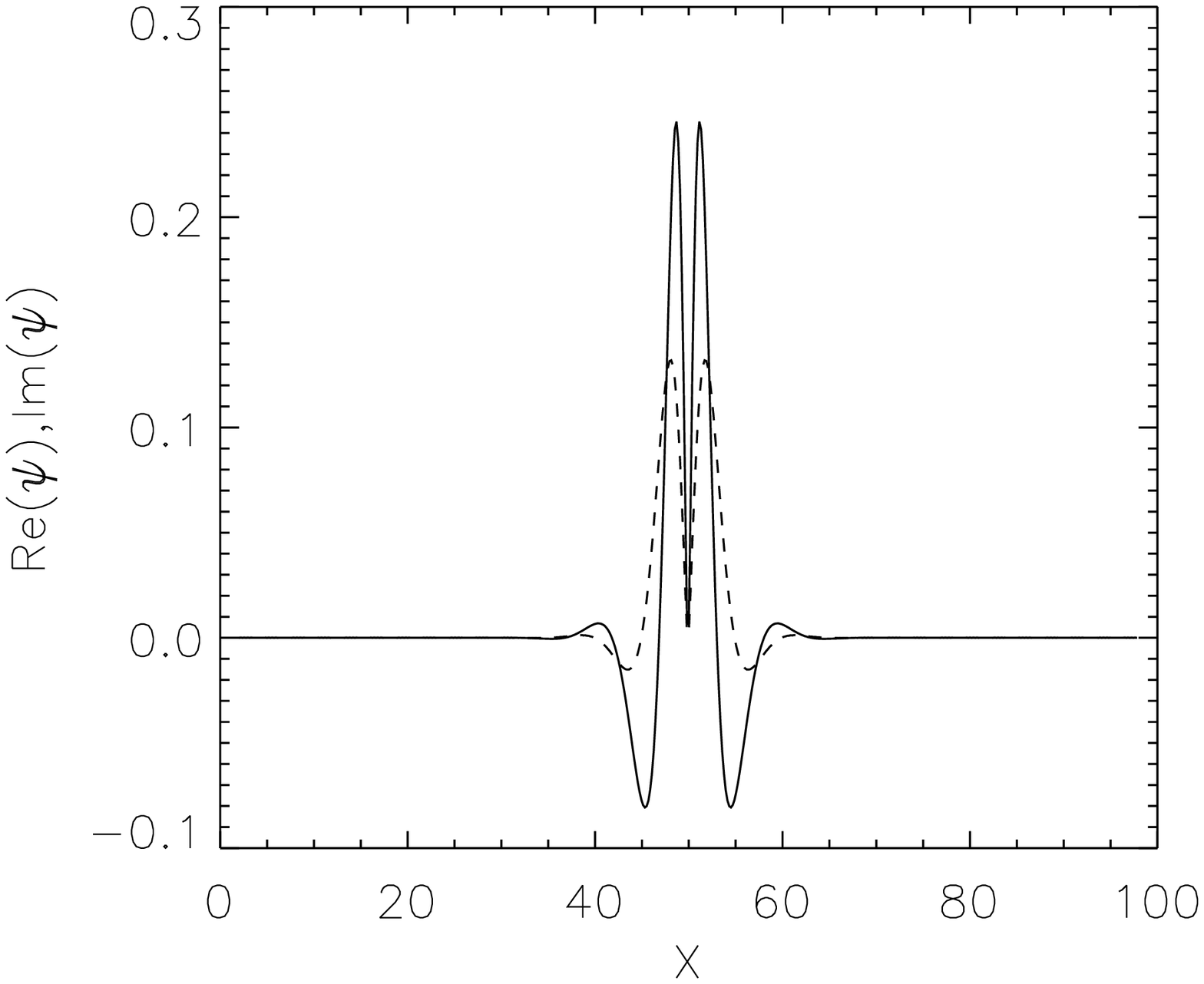}
\caption{Transverse cut of the most stable (left) and unstable (right) eigenmodes of the linear
spectrum shown in Fig.\ref{spectrum}. The solid (dashed) line indicates the real
(imaginary) part of the eigenmode.}
\label{localizedmodes}
\end{figure}

For $\theta > \theta_{\rm SNSL}$ the scenario is as described in Section \ref{sec:snic}, namely
the middle and lower branches coalesce in a SNIC bifurcation. The
main feature of this bifurcation is that it occurs {\it on\/} the limit
cycle, leading to the excitable behavior of region IV. This scenario can be confirmed using
the mode projection technique described in the Appendix of Ref.~\cite{hompump}, that
allows to obtain in quantitative form the phase space of an extended system (described,
e.g., by a PDE), that strictly has an infinite dimension, but whose relevant dynamics is
low-dimensional. In this case, like in the
case of a saddle-loop bifurcation studied in Ref.~\cite{hompump}, we will argue that there
are just two modes that are relevant, at least in the region close to the SNSL
codimension-$2$ point. Fig.~\ref{spectrum} shows the spectrum of eigenvalues (linear
stability analysis for the fundamental solution) at the SNIC bifurcation and close to the SNSL.
The spectrum has a continuous part with eigenvalues lying along the line ${\rm Re}(\lambda)=-1$,
and also a discrete part which is symmetric with the respect to this line. It turns out
that there are only two eigenmodes which are localized in space,
while all the other eigenmodes are spatially extended. The two localized eigenmodes
correspond to the most stable mode and to the one that becomes unstable. These two modes, shown in  Fig.~\ref{localizedmodes}, are
the only relevant for the dynamics of the CS close to the stable fixed point, since the projection of a localized solution onto any of the extended modes is negligible.

\begin{figure}
\includegraphics[width=0.4\textwidth]{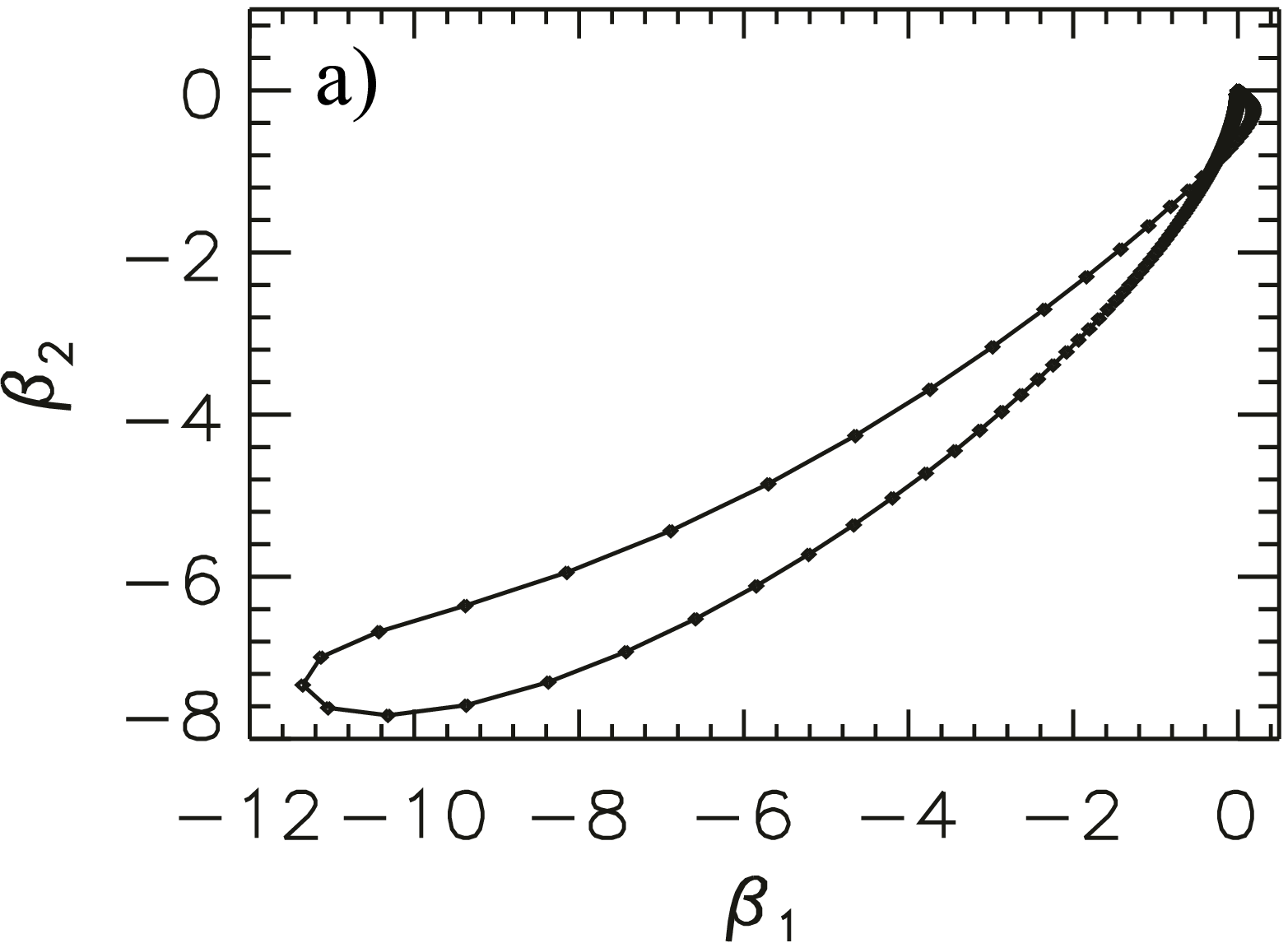}
\includegraphics[width=0.5\textwidth]{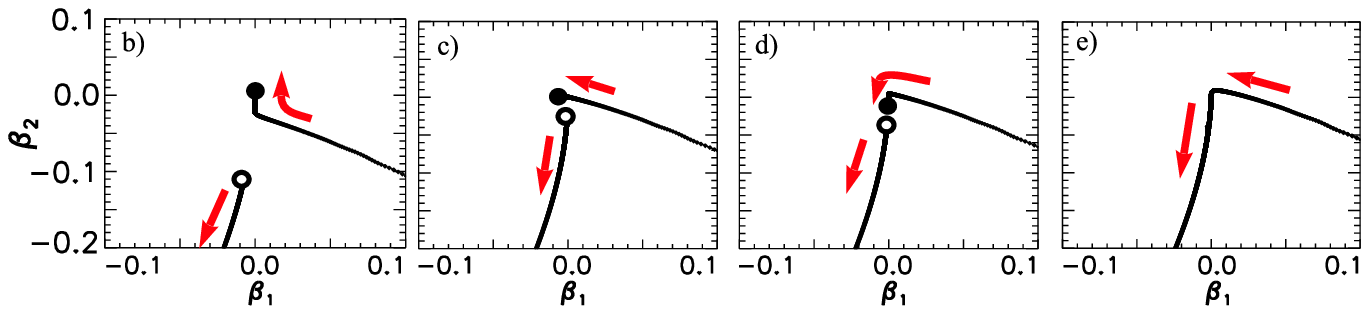}
\caption{a) Full trajectory in the phase space close to the SNIC ($I_s=0.863$). The smaller panels show a zoom of
the region in the phase space close to the fixed point. b) $I_s=0.863$, c) $I_s=0.8634575$, d) $I_s=0.8635$, and e) $I_s=0.864$. Here $I_{sh}=0.3$ and $\theta=1.34$.}
\label{phase_snic}
\end{figure}

Fig.~\ref{phase_snic} shows a quantitative reconstruction of the phase space.
$\beta_1$ ($\beta_2$) corresponds to the amplitude of the projection of the trajectory along the unstable (most stable) eigenmode. Panel a) represents a trajectory in the excitable region IV, close to the SNSL and below the pseudo-bifurcation line.
Panel b) shows a zoom of the trajectory close to the saddle (open circle) and the stable fundamental solution (filled circle). The excitable trajectory departs form the saddle and after long excursion in phase space arrives to the fundamental solution from below (that is, from the side of the saddle). Notice that while close to the fixed points
the dynamics fits very well in a 2-dimensional picture, away from them the lines crosses, indicating that the full
CS dynamics in phase space is not confined to a plane.
Panel c) corresponds to the pseudo-bifurcation, so that the trajectory arrives at the fundamental solution along the most stable direction. Panel d) is just after the pseudo-bifurcation with the trajectory arriving from the other side.
Finally panel e) corresponds to parameters in the oscillatory region V just after the SNIC bifurcation.
Notice that the pseudo-bifurcation line is very close to the SNIC bifurcation since we have taken parameters close to the SNSL and at the SNSL both lines originate tangentially.
As expected, the quantitative picture agrees with the qualitatively picture as one crosses the
SNIC bifurcation in the right panels of Fig.~\ref{sketch_snic}.

For $\theta<\theta_{SNSL}$ the scenario is quite different, as the middle and lower branches
now coalesce off the limit cycle, SN2, what implies that the behavior of the system is oscillatory
on both sides of the SN2 line, not excitable \footnote{Due to the bending of the SL line when
approaching the SNSL there is a region
in which in vertical paths in parameter space crosses the SL line twice. Qualitatively there is
no difference between crossing twice or none the SL line, as one is in the oscillatory bistable
region III when approaches the SN2 line.}
and instead one crosses the SN2 line. This means that one is in a bistable
oscillatory regime, and in crossing the SN2 line a saddle-node off the invariant cycle bifurcation
occurs. SN2 involves the middle and fundamental branches (while SN involves the upper and middle branches).
So, in the end this nothing else that another way of entering region V, of monostable oscillatory
behavior, but instead of reconstructing the limit cycle, here the fundamental stationary solution is
destroyed.

Therefore, and having in mind the behavior for $\theta\rightarrow \infty$ discussed in previous work for homogeneous pump \cite{damia05,hompump}, the overall scenario depicted in Figs.~\ref{phasediag_0.3}-\ref{phasediag_0.7} is organized by three codimension-2 points: a Cusp point, from which two saddle-node bifurcations emerge (SN and SN2);
a SNSL point from which a SNIC line emerges, and that organizes the SN2 and SL lines around; and a Takens-Bodgdanov point, occurring apparently at infinite detuning, where the SN, Hopf and SL are tangent, and that can be seen as the birth of both the Hopf and SL lines. The Takens-Bogdanov point was numerically shown to be present in the
homogeneous case \cite{damia05,hompump}. It is reassuring that the saddle-loop bifurcation line
connects two of the codimension-$2$ points, as it is not in some sense generic,
\footnote{Because it implies that the a limit cycle has a tangency
simultaneously with the stable and unstable manifolds of a saddle point.} and one
does not expect that it emerges {\it out of the blue sky}.
The scenario composed by these three codimension-2 bifurcations has been reported in other systems \cite{Mercader}, and from a theoretical point of view, it can be shown that it appears in the unfolding in 2-dimensional parameter space of a codimension-3 degenerate Takens-Bogdanov point \cite{Dumortier}.

In the limit of homogeneous pump the SN2 and SNIC
lines approach the MI line at $I_s=1$, which, therefore, also contains the Cusp and the SNSL codimension-2 points.
At the Cusp the SN (responsible for the existence of CS) originates while at the SNSL the SL (originated at the Takens-Bogdanov and responsible of the CS excitability observed for homogeneous pump) ends.
Notice that for homogeneous pump close to $I_s=1$ azimuthal instabilities renders the CS unstable to a pattern so it is difficult to study the SN and specially the SL lines in that region. Using a localized pump and then taking the limit to homogeneous pump circumvents these limitations.

\section{Conclusions} \label{sec:conclu}

We have presented a detailed study of the instabilities of solitons in nonlinear Kerr
cavities under the application of a localized Gaussian beam.
Since experimentally CS are typically switched on by applying a transient addressing beam the situation discussed
here can be realized just applying it on a permanent basis.
The CS are sustained by a balance between nonlinearity and dissipation, as in the
case of a homogeneous pump, although now these effects interact with
the localized pump. The localized pump helps to spatially fixed the CS and to control some
of its dynamical properties. After the saddle-node bifurcation that creates the CS,
it starts oscillating and overall exhibit a plethora of bifurcations that are shown
to be organized by three codimension-$2$ points: a Takens-Bogdanov point, (which is also present
for homogeneous pump as discussed already in \cite{damia05,hompump}), a Cusp, and a Saddle-Node Separatrix Loop
(SNSL) points. In this scenario a saddle-loop bifurcation connects the Takens-Bogdanov and the SNSL and the Cusp is connected to the other codimension-2 bifurcations by two saddle-node lines. A line of SNIC bifurcations originates at the SNSL while at the Takens-Bogdanov a Hopf bifurcation line meets tangentially a saddle-node and the saddle-loop lines.

The simultaneously presence in the system of two bifurcations that are
associated to excitable behavior (saddle-loop and SNIC) enriches and completes the picture discussed
for the case of a homogeneous pump. In fact the region for which excitable
behavior is reported, in which the only attractor in the system is the
fundamental solution, leads to two different Class I behaviors (starting at infinite period).
In the excitable region one goes smoothly from no threshold at the onset of the SNIC line to
a finite threshold at the onset of the saddle-loop line.
In fact, the excitable behavior mediated by a SNIC, and reported in this work,
should be easier to observe both numerically and experimentally, and present
some practical features that make it more suitable for practical applications.
In particular, the excitable threshold can be controlled by the intensity of the
localized beam. With an array of properly engineered beams, created for
instance with a spatial light modulator, one could create reconfigurable arrays
of coupled excitable units to all-optically process information. Work in this
direction will be reported elsewhere.

\begin{acknowledgments}
We acknowledge financial support from MEC (Spain) and FEDER (EU)
through Grants FIS2007-60327 (FISICOS) and
TEC2006-10009 (PhoDeCC), and from Govern Balear through Grant PROGECIB-5A (QULMI).
We are grateful to Diego Paz\'o for useful discussions.

\end{acknowledgments}


\end{document}